\documentclass[aps,superscriptaddress,nofootinbib,
floatfix,amsfonts,twocolumn]{revtex4-2}
\usepackage[bookmarksnumbered,bookmarksopen,colorlinks,citecolor=red,linkcolor=blue]{hyperref}
\usepackage{graphicx}
\usepackage{bm}
\usepackage{amsmath}
\usepackage{xcolor}
\usepackage[normalem]{ulem}
\usepackage{booktabs}
\usepackage{multirow}
\usepackage{diagbox}
\usepackage{here}

\graphicspath{{./figs/}}

\begin{document}
\title{$B_c$ enhancement with non-thermalized bottom quarks in nuclear collisions at Large Hadron Collider}
\author{Jiaxing Zhao}
\affiliation{Physics Department, Tsinghua University, Beijing 100084, China\\}
\affiliation{SUBATECH, Universit\'e de Nantes, IMT Atlantique, IN2P3/CNRS, 4 rue Alfred Kastler, 44307 Nantes cedex 3, France}
\affiliation{Helmholtz Research Academy Hesse for FAIR (HFHF), GSI Helmholtz Center for Heavy Ion Physics, Campus Frankfurt, 60438 Frankfurt, Germany}
\affiliation{Institut f\"ur Theoretische Physik, Johann Wolfgang Goethe-Universit\"at,Max-von-Laue-Straße 1, D-60438 Frankfurt am Main, Germany}
\author{Pengfei Zhuang}
\affiliation{Physics Department, Tsinghua University, Beijing 100084, China\\}
\date{\today}

\begin{abstract}
We study $B_c$ production in high-energy nuclear collisions in a transport approach with dissociation and regeneration at finite temperatures. Due to the rare production in p+p collisions and the strong combination of uncorrelated $c$ and $\bar b$ quarks in the quark-gluon plasma, the $B_c$ yield is significantly enhanced in the nuclear collisions at the Large Hadron Collider. 
Moreover, the centrality and momentum-dependent yield of $B_c$ sensitively reflect the thermalization degree of bottom quarks. And the newly observed experimental data favors a far from the thermal bottom quark distribution in the quark-gluon plasma.
\end{abstract}

\maketitle
$B_c$ meson was firstly observed by the CDF Collaboration in 1998~\cite{CDF:1998axz} and then confirmed by the D0~\cite{D0:2008bqs}, CMS~\cite{CMS:2019uhm} and LHCb~\cite{LHCb:2012ihf,LHCb:2013kwl,LHCb:2014mvo,LHCb:2019bem} Collaborations. Theoretically, the mass spectrum of the $c\bar b$ bound state is widely studied by first-principle-based lattice QCD simulations~\cite{Gregory:2009hq,Dowdall:2012ab,Mathur:2018epb}, effective field theories~\cite{Brambilla:2000db,Penin:2004xi,Wang:2012kw,Peset:2018ria,Chang:2019eob,Wang:2022cxy} and non-relativistic potential models~\cite{Gershtein:1994dxw,Zeng:1994vj,Fulcher:1998ka,Eichten:2019gig,Li:2019tbn,Ebert:2002pp,Godfrey:2004ya,Zhao:2020jqu}. Considering the fact that producing a $B_c$ meson requires two pairs of different heavy quarks ($c\bar c$ and $b\bar b$) created in one event, its yield in elementary (hadronic) collisions is very low, and only the ground state and first excited state of $B_c$ meson are experimentally discovered~\cite{ParticleDataGroup:2020ssz}.

It is widely accepted that high-energy nuclear collisions can generate a new state of matter called quark-gluon plasma (QGP) at high temperature. In such nucleus-nucleus collisions, there are two sources for $B_c$ production: One is the initial production which is basically a superposition of the elementary nucleon-nucleon collisions, and the other is the combination of two heavy quarks distributed in the QGP. While the initial production is still weak, the two heavy quarks which combine into a $B_c$ meson are not necessarily from the same elementary collision, and the combination of the off-diagonal heavy quarks from different elemental collisions may lead to a large yield~\cite{Schroedter:2000ek,Liu:2012tn,Chen:2018obq}. This strong $B_c$ enhancement in nuclear collisions, in comparison with hadronic collisions, provides a way to discover higher excited states of $B_c$ meson and a clean signal of the QGP creation in nuclear collisions. Recently, the CMS Collaboration claims the first $B_c$ observation in Pb+Pb collisions~\cite{CMS:2022sxl}. Compared with $J/\psi$ and $\Upsilon$, the nuclear modification factor $R_{AA}$ for $B_c$ is much larger. The related theoretical studies are followed by Refs.~\cite{Chen:2021uar,Wu:2023djn}.

In this paper, we study $B_c$ production and heavy quark thermalization in nuclear collisions at the Large Hadron Collider (LHC). We first consider $B_c$ properties in the vacuum and at finite temperature in a potential model and $B_c$ production in elementary p+p collisions, and then focus on $B_c$ evaluation in nuclear collisions in a transport approach including dissociation and regeneration processes in the QGP.

\begin{figure}[!tb]
\includegraphics[width=0.4\textwidth]{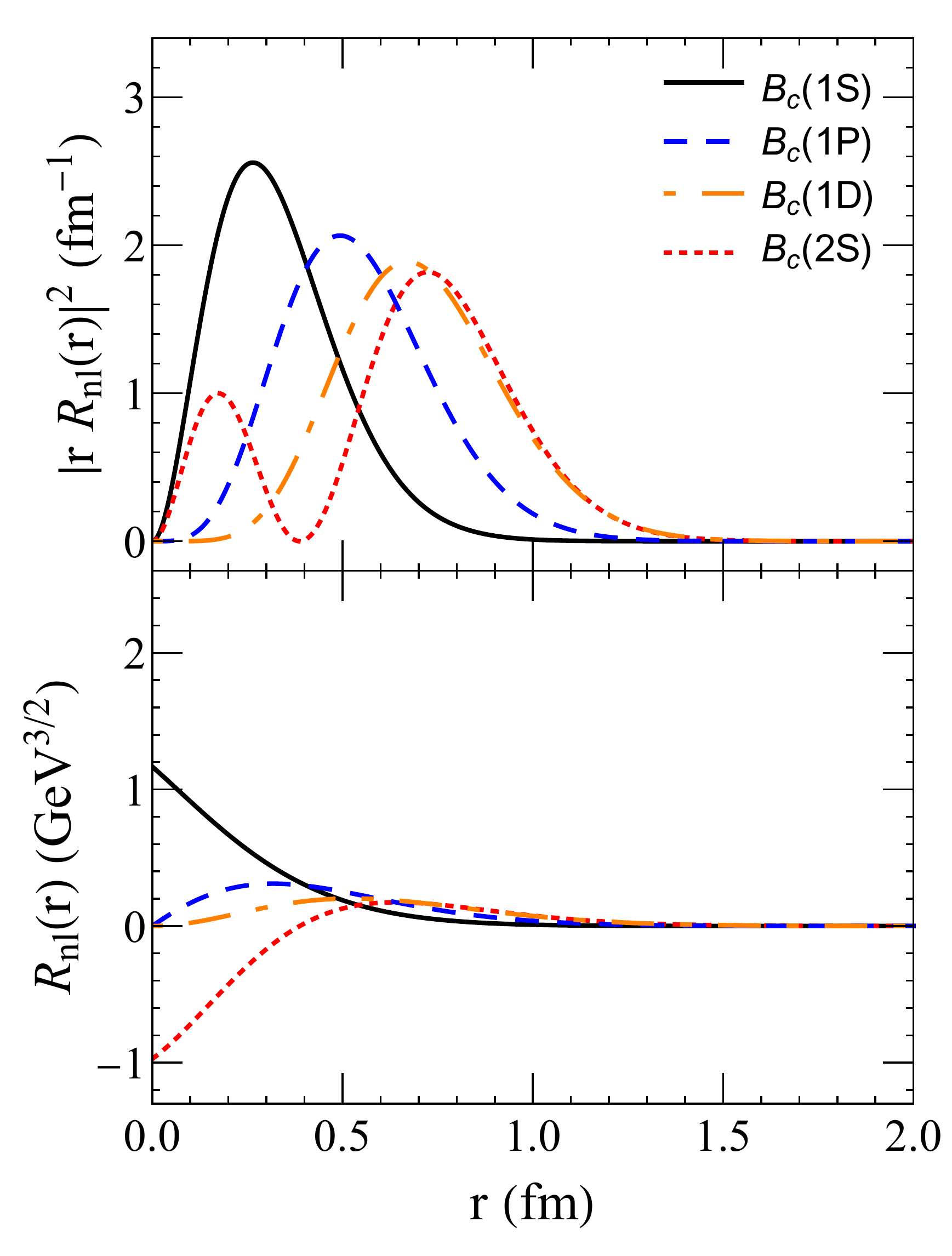}
\vspace{-0.5cm}
\caption{The wavefunctions of the four lowest $B_c$ states, which are below the $B+D$ threshold. }
\label{fig.wf}
\end{figure}

Since charm and bottom quarks are so heavy, the relative motion of the two-body system $c\bar b$ (or $b\bar c$) can be described by the non-relativistic Schr\"odinger equation. Due to the radial symmetry of the interaction between the two quarks, the relative wave function can be separated into a radial part and an angular part,  $\Psi_{nlm}({\bf r})=R_{nl}(r)Y_{lm}(\theta,\phi)$, where $Y_{lm}$ are spherical harmonic functions with quantum numbers $l$ and $m$ describing the orbital angular momentum, and $R_{nl}$ is controlled by the radial equation
\begin{eqnarray}
\label{radial}
&&\left[{1\over 2\mu} \left(-{d^2\over d^2r}-{2\over r}{d\over dr}+{l(l+1)\over r^2} \right)+V(T,r) \right]R_{nl}(r)\nonumber\\
&=& E_{nl} R_{nl}(r)
\end{eqnarray}
with the reduced mass $\mu=m_b m_c/(m_b+m_c)$ and relative energy $E_{nl}$. The interaction potential $V(T,r)$ becomes complex at finite temperature, the real part is governed by the color screening mechanism~\cite{Brambilla:2008cx}, and the latter comes from the gluo-dissociation~\cite{Brambilla:2011sg} and inelastic collisions~\cite{Brambilla:2013dpa}. Neglecting the imaginary part as a first approximation, and introducing a Debye mass $m_D$~\cite{Laine:2006ns} to describe the strength of the color screening, the in-medium potential between a heavy quark and an anti-heavy quark can be expressed as~\cite{Lafferty:2019jpr},
\begin{eqnarray}
\label{potential}
V(T,r) &=& -\alpha\left[m_D+e^{-m_Dr}/r\right]\nonumber\\
&& +\sigma/m_D\left[2-(2+m_Dr)e^{-m_Dr}\right],
\end{eqnarray}
where the temperature dependence of the Debye mass $m_D(T)$ can be extracted from fitting the lattice simulated potential~\cite{Burnier:2014ssa,Burnier:2015tda}. In the zero temperature limit $T\to 0$, the screening mass disappears $m_D(0)\to 0$, the potential is reduced to the Cornell form in vacuum $V(0,r)=-\alpha/r+\sigma r$. The quark masses and potential parameters in the radial equation (\ref{radial}) can be fixed by fitting the known charmonium and bottomonium masses which leads to $m_c=1.29$ GeV, $m_b=4.7$ GeV, $\alpha=0.4105$ and $\sigma=0.2$ GeV$^2$~\cite{Zhao:2020jqu}. The $B_c$ mass $M_{nl}$ is defined as $M_{nl}=m_c+m_b+E_{nl}$. By solving the radial equation in vacuum, the calculated masses below the $B$ + $D$ threshold and the comparison with the experimental data are shown in Table \ref{table1}. The wavefunctions are shown in Fig.~\ref{fig.wf}. 
\begin{table}
	\renewcommand\arraystretch{1.8}
	\setlength{\tabcolsep}{2.5mm}
	\begin{tabular}{c|c|c|c|c}
		\toprule[1pt]\toprule[1pt]
		\multicolumn{1}{c|}{} & \multicolumn{1}{c|}{$1S$} &   \multicolumn{1}{c|}{$1P$} &   \multicolumn{1}{c|}{$1D$} &   \multicolumn{1}{c}{$2S$} \tabularnewline
		\midrule[1pt]
		$\text{Exp.}$ & 6.275 & -  & - & 6.872 \tabularnewline
		\bottomrule[1pt]
		$\text{Theo.}$ & 6.332 & 6.735 &  7.017 & 6.908 \tabularnewline
		\bottomrule[1pt]
	\end{tabular}
	\caption{The experimentally measured~\cite{ParticleDataGroup:2020ssz} and model calculated $B_c$ meson masses $M_{nl}$ in the unit of GeV.}
	\label{table1}
\end{table}

\begin{figure}[!tb]
\includegraphics[width=0.4\textwidth]{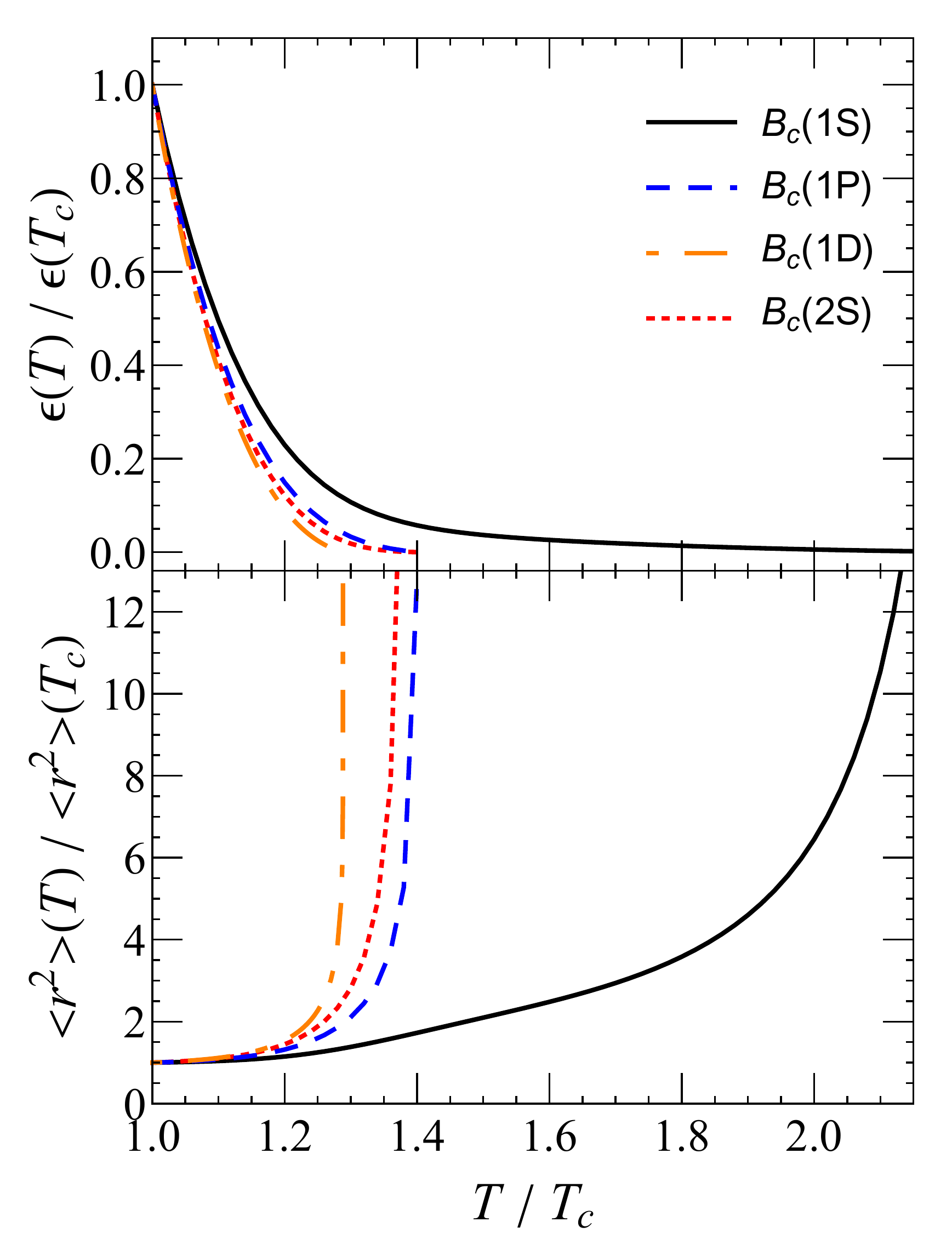}
\vspace{-0.5cm}
\caption{The temperature dependence of the binding energy and averaged radius square for the four lowest $B_c$ states, scaled by their values at the deconfinement temperature $T_c=170$ MeV. }
\label{fig.be}
\end{figure}

The long-ragne color interaction between two heavy quarks is screened by the surrounding quarks and gluons in the hot medium, and the potential becomes saturated at distance $r>r_D\sim 1/m_D$ with the value $V(T,\infty)=-\alpha m_D+2\sigma/m_D$. When the screening length $r_D$ is less than the typical size of the $B_c$ meson, it is melted by the medium. The melting temperature $T_D$ can be defined through the vanishing binding energy $\epsilon_{nl}(T_D)=E_{nl}(T_D)-V(T_D,\infty)=0$, corresponding to the infinite averaged radius $\langle r \rangle (T_D) \to \infty$. The temperature dependence of the binding energy and averaged radius square, scaled by their values at the deconfinement temperature $T_c$, is shown in Fig.~\ref{fig.be}. The melting temperature is $T_D\approx 2.1 T_c$ for $B_c(1S)$ which is comparable with the $J/\psi$ melting temperature~\cite{Lafferty:2019jpr} and $T_D\approx 1.3 T_c$ for the excited states $B_c(1P)$, $B_c(1D)$, and $B_c(2S)$.

The study on heavy flavor hadrons in nuclear collisions needs not only their properties in hot medium but also their production in elementary nucleon-nucleon collisions. In elementary collisions at high energies, $B_c$ meson can be generated via gluon-gluon fusion $gg\to B_c + b + \bar c$ and quark-antiquark annihilation $q\bar q\to B_c + b + \bar c$, and the former is the dominant one~\cite{Chang:2003cq}. While the $c\bar c$ and $b\bar b$ pair production can be calculated in the frame of perturbative QCD (pQCD), since the heavy quark masses are larger than the QCD scale, $m_c,m_b\gg \Lambda_{QCD}$, the combination of $c$ and $\bar b$ quarks into a $B_c$ is a non-perturbative process. In the factorization picture, the non-perturbative factor is related to the non-relativistic wave function or its first non-vanishing derivative~\cite{Brambilla:2021abf} of the $B_c$ state at the origin. The $B_c$ production cross section at LHC energy can be calculated by using the well-developed generator BCVEGPY2.2~\cite{Chang:2003cq,Chang:2015qea} which is now extended to include the 3P state. The generator has been used to predict the $B_c$ yield in p+p, p+$\bar {\text p}$, e$^+$+e$^-$ and even heavy-ion collisions~\cite{LHCb:2014mvo,Zheng:2015ixa,Chen:2018obq}.
\begin{figure}[!tb]
	\includegraphics[width=0.4\textwidth]{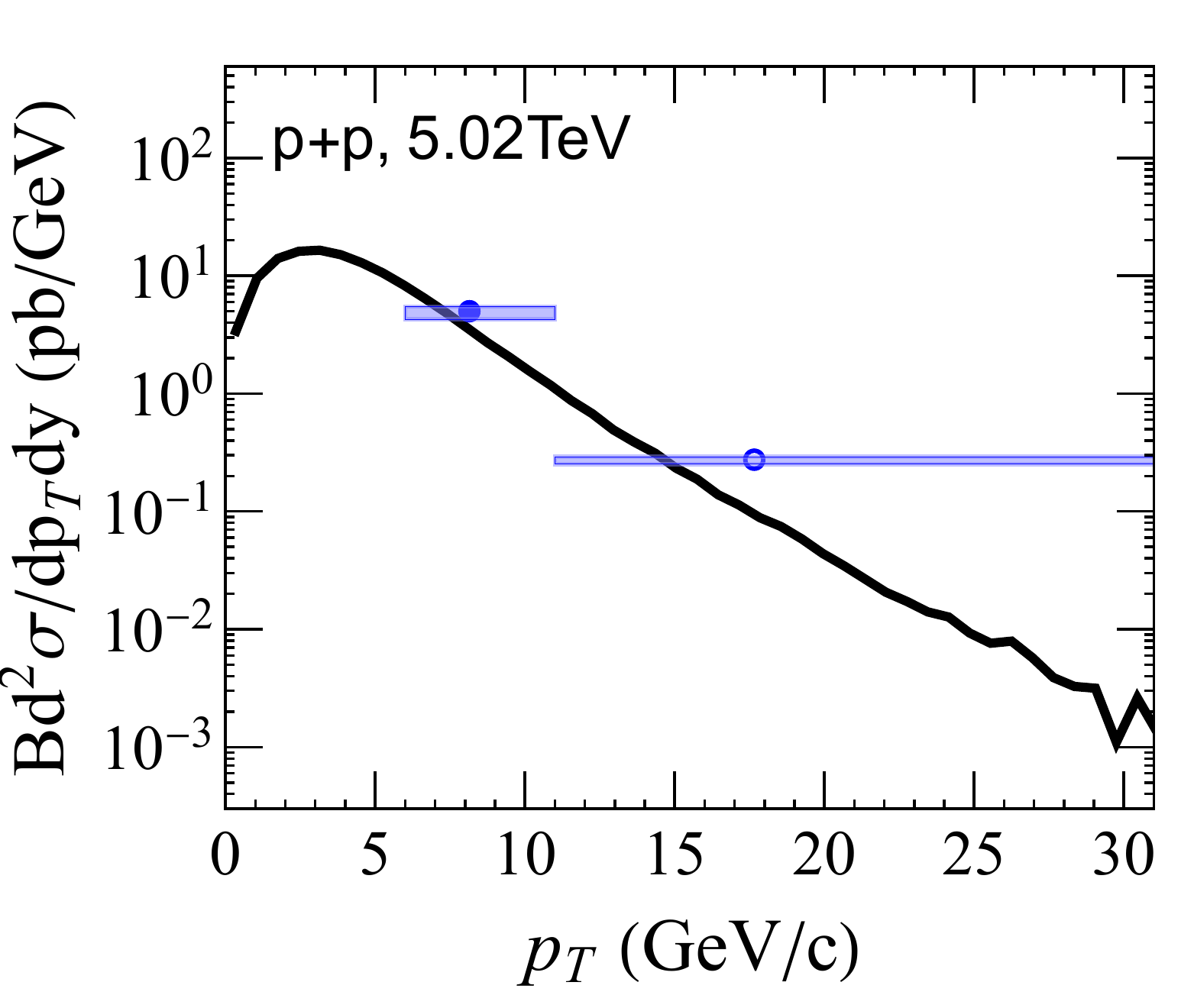}\\
        \includegraphics[width=0.4\textwidth]{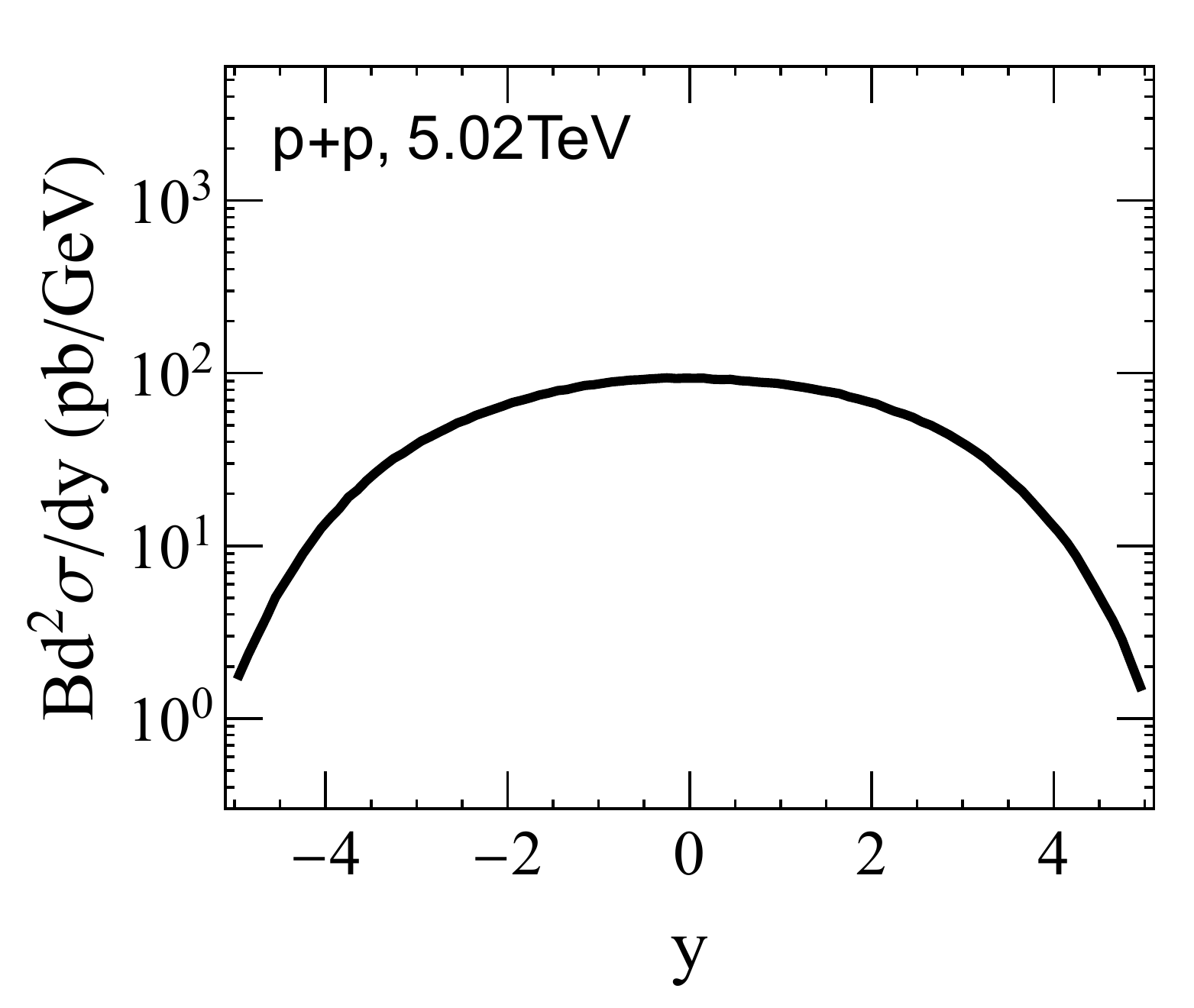}
    \vspace{-0.5cm}
	\caption{The transverse momentum (upper panel) and rapidity (lower panel) dependence of the $B_c$ production cross sections p+p collisions at $\sqrt{s}=5.02~\rm TeV$. The results are calculated by the event generator BCVEGPY2.2. The experimental data in the upper panel are from the CMS Collaboration~\cite{CMS:2022sxl}.}
	\label{fig.bcpp}
\end{figure}

The wave function or its first non-vanishing derivative for different $B_c$ state at the origin is determined by the Schr\"odinger equation (\ref{radial}) with the solution, as shwon in the bottom panel of Fig.~\ref{fig.wf},
\begin{eqnarray}
\label{zero}
R_{00}(0) &=& 1.163\ \text {GeV}^{3/2},  \nonumber\\
R'_{01}(0) &=& 0.382\ \text {GeV}^{5/2}, \nonumber\\
R''_{02}(0) &=& 0.197\ \text {GeV}^{7/2}, \nonumber\\
R_{20}(0) &=& -0.970\ \text {GeV}^{3/2}.
 \end{eqnarray}
Taking them as the input of the generator BCVEGPY2.2, the differential cross sections for $B_c$ mesons in p+p collisions at $\sqrt s=5.02$ TeV are shown in Fig.~\ref{fig.bcpp}. For the $D$-wave state, the production yield is approximately one half of that for the $P$-wave state. For comparison with the experimental data of $B_c^+$, which is the spin singlet state $^1S_0$, the feeddown contributions from all excited states, such as $1P$, $1D$, and $2S$ should be included.
The experimental data is represented by the branching ratio $B(B_c\to (J/\psi\to \mu^+\mu^-)\mu^+v_{\mu})$, where $B(J/\psi\to \mu^+\mu^-)=5.96\%$ for $J/\psi$ is given by Particle Data Group (PDG)~\cite{ParticleDataGroup:2020ssz}. However, the decay branching ratio $B(B_c\to J/\psi\mu^+v_{\mu})$ has not been observed in experiments. There are many theoretical predictions and give an uncertainty range from 1.2\% to 6.7\%~\cite{Ebert:2003cn,Hernandez:2006gt,Qiao:2012vt}. A good explanation of the experimental data in the framework of BCVEGPY2.2 leads to $B(B_c\to J/\psi\mu^+v_{\mu})=6\%$, which is almost the upper limit. The total transverse momentum distribution is shown in the upper panel of Fig.~\ref{fig.bcpp} and the lower panel is the rapidity distribution. By integrating out the transverse momentum, the total cross section (with feeddown contributions) is is $\sigma^{B_c(1S)}_{pp} \approx 140$ nb, and the total cross section per unit of rapidity is $d\sigma^{B_c(1S)}_{pp}/dy \approx 26$ nb in the central rapidity region $|y|<2$.

We now turn to the calculation of $B_c$ production in nuclear collisions. We consider first the evolution of the hot medium which is the background of the $B_c$ motion. The created quark-gluon matter is expected to reach local equilibrium at about $\tau_0\approx 0.6$ fm/c~\cite{Schenke:2010nt} after the nuclear collision, then its space-time evolution is controlled by the conservation laws of the system, namely the hydrodynamic equations. In this paper, we employ a (2+1)-dimensional hydrodynamic model, the MUSIC package~\cite{Schenke:2010nt,McDonald:2016vlt}, to characterize the space and time dependence of the temperature and velocity of the hot medium. To close the hydrodynamic equations, an equation of state for both the QGP and hadron phases is needed. We adopt the model ``s95p-v'' which matches the Lattice QCD data at high temperature and the hadron resonance gas at low temperature~\cite{Huovinen:2009yb}. The two phases are connected with a smooth crossover at temperature $T_c=170$ MeV. An effective shear viscosity $\eta/s=0.08$~\cite{Bernhard:2016tnd} and a zero bulk viscosity is chosen in the calculation.

Since the melting temperature $T_D$ is larger than the deconfinement temperature $T_c$, the survived $B_c$ mesons can not be thermalized in the QGP medium through color interaction. The distribution function of $B_c$ mesons produced in the high energy nuclear collisions with impact parameter $b$, $f_{B_c}(p,x|b)$, is controlled by the Boltzmann transport equation~\cite{Yan:2006ve},
\begin{eqnarray}
\label{transport}
&&\left[\cosh(y-\eta){\partial\over\partial\tau}+{\sinh(y-\eta)\over\tau}{\partial\over\partial\eta}+{\bm v}_T\cdot{\bm \nabla}_T\right]f_{B_c} \nonumber\\
&=& -\alpha f_{B_c} +\beta,
\end{eqnarray}
where $\eta=1/2\ln[(t+z)/(t-z)]$ and $y=1/2\ln[(E+p_z)/(E-p_z)]$ are $B_c$ space-time rapidity and momentum rapidity, and ${\bm v}_T={\bm p}_T/E_T$ is the transverse velocity with transverse energy $E_T=\sqrt{m_{B_c}^2+{\bm p}_T^2}$. The second and third terms on the left hand side arise from the free streaming of $B_c$, leading to the leakage effect in the longitudinal and transverse directions. The $B_c$ suppression and regeneration in the QGP are reflected in the loss term $\alpha$ and gain term $\beta$. Considering only the gluon dissociation process for the loss term and its inverse process for the gain term, $\alpha$ and $\beta$ are expressed as~\cite{Yan:2006ve}
\begin{eqnarray}
\label{alphabeta}
\alpha&=&{1\over 2E_T}\int{d^3{\bm p}_g \over(2\pi)^32E_g}W_{gB_c}^{{\bar b}c}(s)f_g({p}_g,x), \nonumber\\
\beta&=&{1\over 2E_T}\int {d^3{\bm p}_g \over(2\pi)^32E_g}{d^3{\bm p}_{\bar b}\over(2\pi)^32E_{\bar b}}{d^3{\bm p}_{c} \over(2\pi)^32E_{c}}  \nonumber\\
&\times& W_{{\bar b}c}^{gB_c}(s)f_{\bar b}({p}_{\bar b},x)f_{c}({p}_{c},x) \nonumber\\
&\times& (2\pi)^4\delta^{(4)}(p+p_g-p_{\bar b}-p_{c}),
\end{eqnarray}
where $f_g$ is the gluon thermal distribution $f_g(p,x)=d_g/(e^{p^\mu u_\mu(x)/T(x)}+1)$ with the degenerate factor $d_g=16$ and the medium temperature $T(x)$ and velocity $u_\mu(x)$ determined by the hydrodynamics. The dissociation probability $W_{gB_c}^{{\bar b}c}$ is related to the cross section $\sigma_{gB_c}^{{\bar b}c}$, and the regeneration probability $W_{\bar bc}^{gB_c}$ can be determined by detail balance~\cite{Yan:2006ve}. In this work, we consider the gluon dissociation process $g+B_c \to \bar b+c$, an analogy to the photon dissociation of electromagnetic bound states. Similar to the charmonium dissociation calculated with operator-production-expansion (OPE) method~\cite{Peskin:1979va,Bhanot:1979vb,Arleo:2001mp}, the cross sections for the 1S, 1P and 2S states of $B_c$ meson read
\begin{eqnarray}
\label{cross1}
\sigma_{1S}(\omega)&=&A_0 (r-1)^{3/2}/r^5,\nonumber\\
\sigma_{1P}(\omega)&=&4A_0 (r-1)^{1/2}(9r^2-20r+12)/r^7,\nonumber\\
\sigma_{2S}(\omega)&=&16A_0 (r-1)^{3/2}(r-3)^2/r^7,
\end{eqnarray}
and an extension of the OPE to the D-wave state is done for the first time in this work and gives 
\begin{equation}
\label{cross2}
\sigma_{1D}(\omega) = 32A_0 (r-1)^{3/2}(21r^2-48r+32)/r^9
\end{equation}
with the gluon energy $\omega$, the ratio $r=\omega/\epsilon$ and the coefficient $A_0=2^{11}\pi/(27\sqrt{(2\mu)^3\epsilon})$. Considering the fact that, at high temperature comparable with the binding energy $\epsilon$, another dissociation named inelastic scattering $q+B_c\to \bar b+c +q$ contributes remarkably~\cite{Brambilla:2008cx}, where $q$ stands for quark, antiquark and gluon. To include this contribution, we take an effective binding energy $\epsilon_{eff}(T)=0.8\epsilon(T=0)$ to replace the binding energy shown in Fig.~\ref{fig.be}, which gives a similar decay width as the lattice simulations~\cite{Burnier:2014ssa,Burnier:2015tda}. Furthermore, when the medium temperature is above the dissociation temperature $T_D$, the bound state is strongly screened, we suspend the calculation and simply take $\alpha=\beta=0$.

Due to the energy loss in the hot medium, the heavy quark distribution is in principle between the pQCD and equilibrium distributions. From the experimental data at the Relativistic Heavy Ion Collider (RHIC) and LHC~\cite{STAR:2017kkh,ALICE:2020pvw}, the observed large quench factor and elliptic flow for charmed mesons indicate a strong interaction between charm quarks and the medium. One can then take a kinetically equilibrated distribution for charm quarks,
\begin{equation}
\label{thermal}
f_c(p,x)=\rho_c(x)N(x)/(e^{p^\mu u_\mu(x)/T(x)}+1),
\end{equation}
where $N(x)$ is the normalization factor. The number density $\rho_c(x)$ is controlled by the charm conservation equation, $\partial_\mu[\rho_c(x)u^\mu(x)]=0$. At the LHC energy, charm quarks are produced mainly via initial binary collisions, and the initial density is governed by the nuclear geometry,
\begin{equation}
\label{geometry}
\rho_c(x_0) = {T_A({\bm x}_T+{\bm b}/2)T_B({\bm x}_T-{\bm b}/2)\cosh\eta\over\tau_0}{d\sigma_\text{pp}^{c\bar c}\over d\eta},
\end{equation}
where $T_A$ and $T_B$ are the thickness functions for the two colliding nuclei~\cite{Miller:2007ri}, $\bm b$ is the impact parameter, and $d\sigma^{c\bar c}_\text{pp}/d\eta=1.165$ mb is the rapidity distribution of the charm quark production cross section in p+p collisions at $\sqrt s=5.02$ TeV, extracted from the new experimental data~\cite{ALICE:2021dhb}.

Bottom quarks are hard to be fully thermalized. In the equilibrium limit, their thermal distribution is the same as that for charm quarks. In another limit, the initially produced bottom quarks do not interact with medium and their distribution satisfies the Vlasov equation,
\begin{equation}
\label{transport2}
p^\mu \partial_\mu f_{\bar b}(p,x)=0.
\end{equation}
Considering the strong Lorentz attraction at LHC energy, the bottom quarks can be assumed to be generated at $z=t=0$, their longitudinal rapidities $y$ and $\eta$ become the same, and the solution of the transport equation is
\begin{equation}
\label{pQCD}
f_{\bar b}(p,x) = \rho_{\bar b}({\bm x}_T - {\bm p}_T/m_b\tau)\delta(\eta-y){(2\pi)^3 \over \tau E_T}{d^3N_{\bar b}\over d^2{\bm p}_T dy}
\end{equation}
with the initial transverse density $\rho_{\bar b}({\bm x}_T) = T_A({\bm x}_T+{\bm b}/2)T_B({\bm x}_T-{\bm b}/2)$. The momentum distribution in the solution can be calculated by  the fixed order next to leading log (FONLL)~\cite{FONLL}.

In real case considering bottom quark interaction with the medium, the relativistic Langevin equation can be used to simulate the evolution of bottom quarks in QGP,
\begin{equation}
{d{\bf p}\over dt}= -\eta {\bf p} +{\bf \xi},
\end{equation}
with $\bf p$ is the momentum of bottom quark. The drag term $\eta$
is connected with the momentum diffusion coefficient via 
the fluctuation-dissipation relation, 
$\eta=\kappa/(2TE_b)$, where the bottom quark energy 
is $E_b=\sqrt{m_b^2+{\bf p}^2}$. 
The momentum diffusion coefficient $\kappa$ is related to the spatial diffusion coefficient $\mathcal{D}_s$ through, $\kappa \mathcal{D}_s = 2T^2$. 
The stochastic term ${\bf \xi}$ is treated as white noise. Neglect the 
momentum dependence in the ${\bf \xi}$, it satisfies the 
relation, 
\begin{align}
\langle \xi^{i}(t)\xi^{j}(t^\prime)\rangle =\kappa \delta ^{ij}\delta(t-t^\prime), 
\end{align}
where the index $i,j$ represents three dimensions. 
With a spatial diffusion coefficient $2\pi TD_s\approx 4$, which has been used in our previous study~\cite{Chen:2017duy} and close to the recent lattice results~\cite{Altenkort:2023eav}, the obtained distribution lies in between the thermal and pQCD distributions (\ref{thermal}) and (\ref{pQCD}). Since the transverse expansion of the QGP medium is much weaker than the longitudinal expansion, the transverse diffusion of the initially produced bottom quarks can be approximately neglected~\cite{Liu:2009nb}. The bottom quark density in coordinate space can then be expressed as
\begin{equation}
\rho_{\bar b}(x) = {T_A({\bm x}_T+{\bm b}/2)T_B({\bm x}_T-{\bm b}/2)\cosh \eta \over \tau}{d\sigma_\text{pp}^{b\bar b}\over d\eta}.
\end{equation}
We take the rapidity distribution of bottom quark production cross section in p+p collisions $d\sigma^{b\bar{b}}_\text{pp}/d\eta=45\mu b$ as suggested by the experimental data~\cite{Andronic:2015wma}.
\begin{figure}[!tb]
\includegraphics[width=0.4\textwidth]{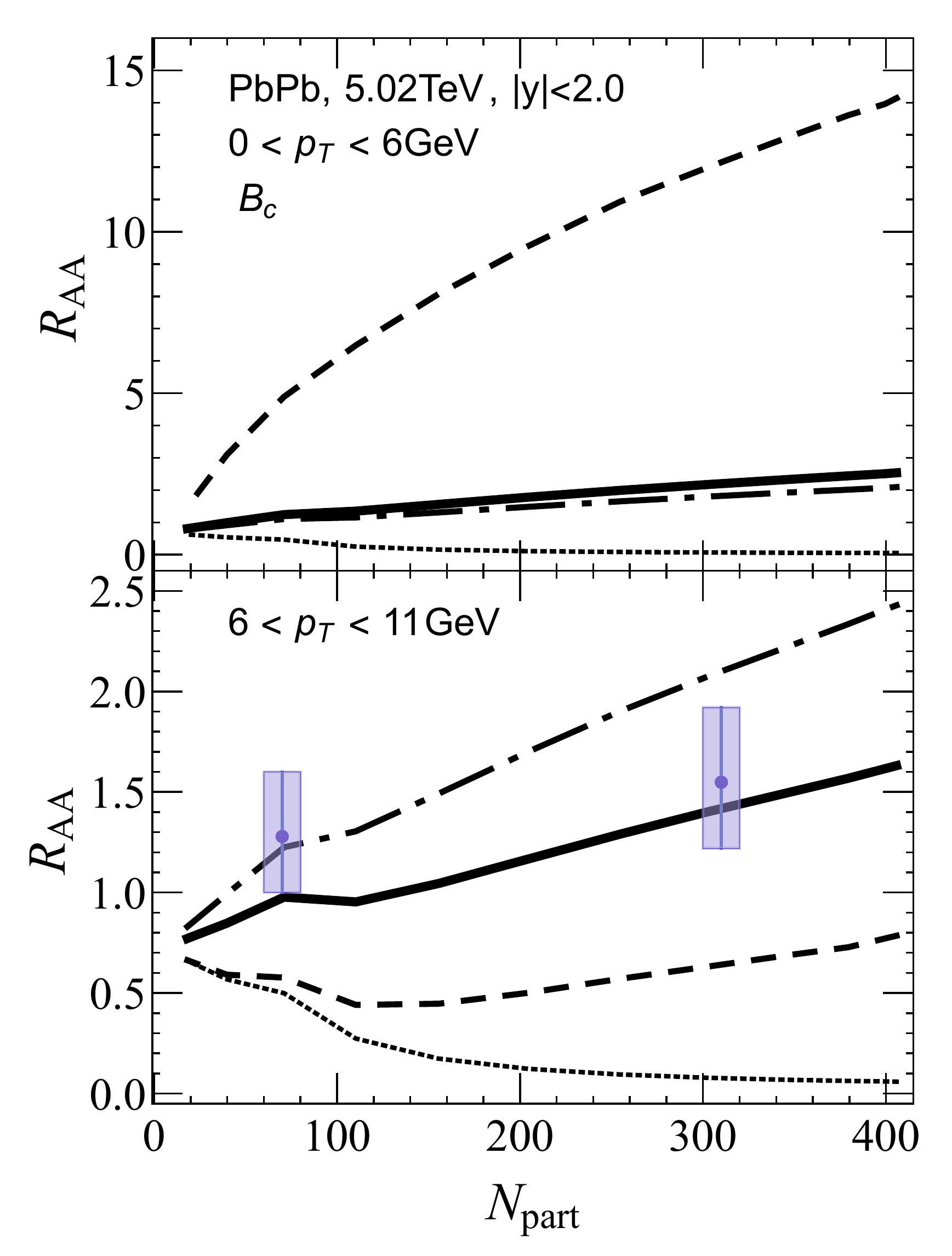}
\caption{The nuclear modification factor $R_{AA}$ for $B_c(1S)$ mesons as a function of the number of participating nucleons $N_\text{part}$ in two transverse momentum regions $0<p_T<6$ GeV (upper panel) and $6<p_T<11$ GeV (lower panel) in Pb+Pb collisions at colliding energy $\sqrt {s_{\rm NN}} =5.02$ TeV. Dotted lines indicate the initial fraction. Dashed, dot-dashed, and solid lines represent the total production with thermal, pQCD, and the Langevin distribution for bottom quarks. The experimental data are from the CMS Collaboration~\cite{CMS:2022sxl}.}
\label{fig.raanpart}
\end{figure}
\begin{figure}[!tb]
\includegraphics[width=0.4\textwidth]{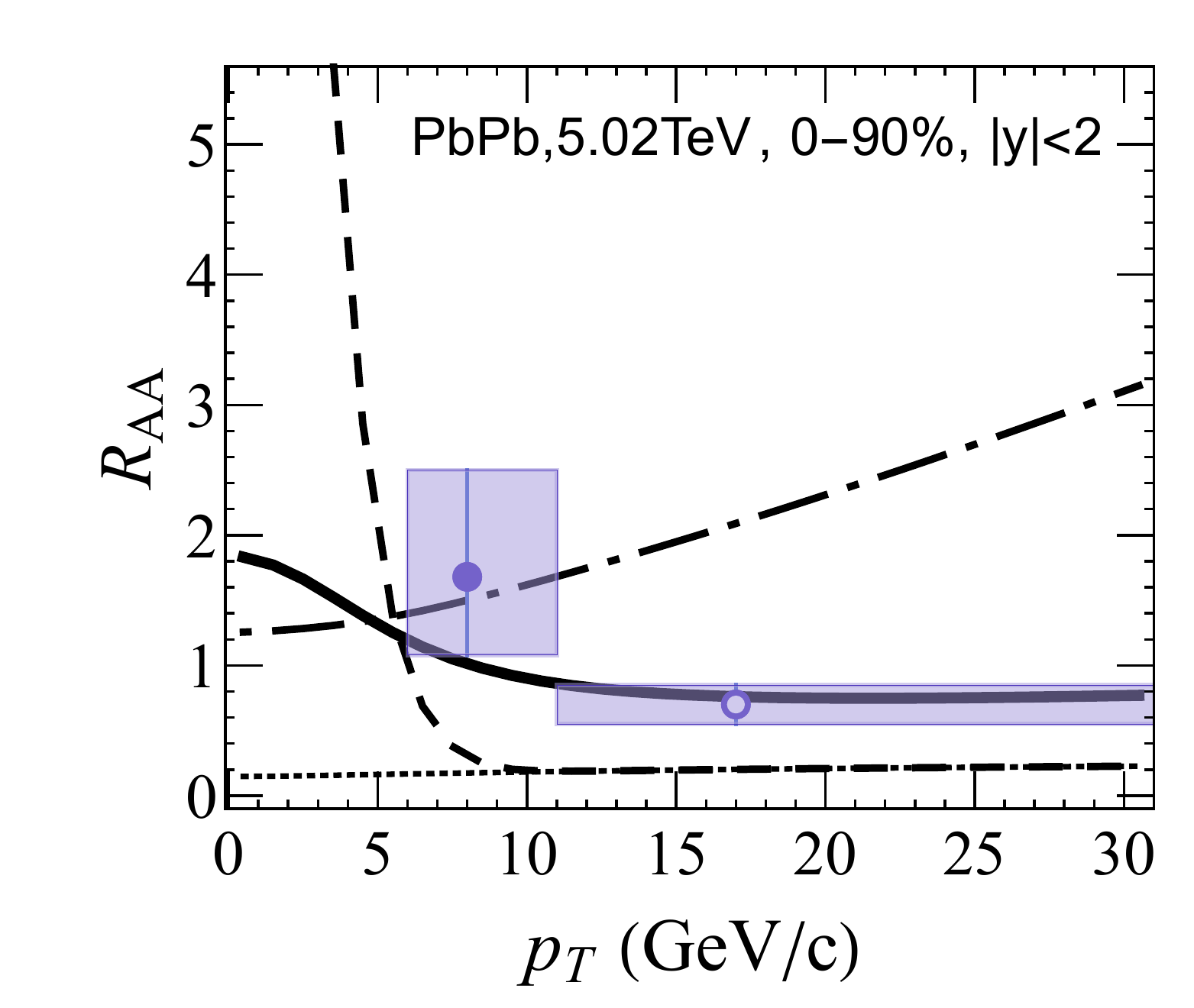}
\caption{The nuclear modification factor $R_{AA}$ for $B_c(1S)$ mesons as a function of transverse momentum $p_T$ in Pb+Pb collisions at centrality $0-90\%$ and colliding energy $\sqrt {s_{\rm NN} }=5.02$ TeV. Dotted lines indicate the initial fraction. Dashed, dot-dashed, and solid lines represent the total production with thermal, pQCD, and the Langevin distribution for bottom quarks. The experimental data are from the CMS Collaboration~\cite{CMS:2022sxl}.}
\label{fig.raapt}
\end{figure}

With the known loss and gain terms, the transport equation (\ref{transport}) can be solved analytically~\cite{Zhou:2014kka}. The only thing left is the initial $B_c$ distribution, which is a geometrical superposition of the corresponding p+p collisions along with modifications from cold nuclear matter effects. The cold effects include usually the nuclear shadowing~\cite{Mueller:1985wy}, Cronin effect~\cite{Cronin:1974zm,Hufner:1988wz} and nuclear absorption~\cite{Gerschel:1988wn}. At the LHC energy, the collision time for two heavy nuclei to pass through each other is much shorter than the $B_c$ formation time, the nuclear absorption can be safely neglected. The Cronin effect broadens the momentum of the initially produced $B_c$ mesons, which can be included through replacing the mean-square transverse momentum $\langle p_T^2\rangle$ in p+p collisions by $\langle p_T^2\rangle+a_{gN}l$~\cite{Huefner:2002tt} in nuclear collisions, where $a_{gN}$ is the Cronin parameter which characters the averaged transverse momentum square obtained from the gluon scattering with a unit of length of nucleons, and $l$ is the mean trajectory length of the two gluons in the two nuclei before the $B_c$ formation. We take $a_{gN}$=0.1 GeV$^2$/fm~\cite{PHENIX:2007tnc}. The shadowing effect changes the parton distribution in a nucleus relative to that in a nucleon, the
modification factor $\mathcal{R}_g$ can be simulated by the EPS09 package~\cite{Helenius:2012wd}. Taking into account the shadowing and Cronin effects, the initial $B_c$ distribution can be expressed as
\begin{eqnarray}
&& f_{B_c}(p,x_0|{\bm b})\nonumber\\
&=& {(2\pi)^3\over \tau_0E_T}\int dz_Adz_B\rho_A({\bm x}_T+{\bm b}/2, z_A)\rho_B({\bm x}_T-{\bm b}/2, z_B)\nonumber\\
&& \times\mathcal{R}_g(x_1,\mu_F,{\bm x}_T+{\bm b}/2)\mathcal{R}_g(x_2,\mu_F,{\bm x}_T-{\bm b}/2)\nonumber\\
&& \times\bar f_{B_c}^\text{pp}(p,x_0|{\bm b}),
\end{eqnarray}
where $\rho_A$ and $\rho_B$ are nucleon distribution functions of the two colliding nuclei which are taken as Woods-Saxon distribution, the longitudinal momentum fractions of the two initial gluons are defined as $x_{1,2}=E_T/\sqrt{s_{\rm NN}}e^{\pm y}$, the factorization scale is taken as $\mu_F=E_T$, and the Cronin-effect modified distribution $\bar f^\text{pp}_{B_c}$ in p+p collision is given by the BCVEGPY2.2 generator.

The centrality and transverse momentum dependence of the nuclear modification factor $R_{AA}$ for $B_c(1S)$ mesons are shown in Figs.~\ref{fig.raanpart} and~\ref{fig.raapt}. Aiming to compare with the newly observed data~\cite{CMS:2022sxl}, both ground and excited states are included. The excited states below the mass threshold of $B$+$D$ decay $100\%$ to the ground state $B_c^+$. Except for the special case of high transverse momentum in peripheral collisions, see the bottom left corner of Fig.~\ref{fig.raanpart}, most of the initially produced $B_c$ mesons in the general case are melted in the hot medium, and the total yield is controlled by the regeneration, independent of the bottom quark distribution in the QGP. In the limit of thermal distribution, the low $p_T$ yield is hugely enhanced, see the upper panel of Fig.~\ref{fig.raanpart} and the left of Fig.~\ref{fig.raapt}, while in the other limit of pQCD distribution, the high $p_T$ yield is largely increased, see the lower panel of Fig.~\ref{fig.raanpart} and the right of Fig.~\ref{fig.raapt}. From the comparison with the experimental data of the CMS Collaboration, bottom quarks are not thermalized in heavy ion collisions at LHC energy, even in very central collisions, and the Langevin distribution considering heavy quark energy loss can describe the bottom quark evolution reasonably well.

In this paper, we studied $B_c$ production in high-energy nuclear collisions at LHC energy. The $B_c$ transport in the created QGP is characterized by a Boltzmann equation with dissociation and regeneration-related loss and gain terms. The initial production via p+p collisions and cold nuclear matter effects are reflected in the initial distribution. The $B_c$ static properties are calculated in a potential model, and the QGP evaluation is governed by hydrodynamic equations. We have two conclusions. 1) Bottom quarks are not thermalized in the QGP. $B_c$ production is sensitive to bottom quark distribution. Different from charm quarks which are believed to be thermalized via strong interaction with the medium, from the comparison with the recent CMS data, bottom quarks in the QGP can be described by a Langevin equation and are far from thermalization. 2) $B_c$ production is enhanced relative to p+p collisions. In comparison with $J/\psi$ and $\Upsilon$ where the nuclear modification factor $R_{AA}$ is usually less than or at most approaches to the unit, the $B_c$ $R_{AA}$ is in general larger than the unit, indicating a $B_c$ enhancement due to the rare production in p+p collisions and strong regeneration in QGP.

{\bf Acknowledgement:}
We thank Dr. Baoyi Chen, Xuchang Zheng for helpful discussions, and Biaogang Wu for pointing out a small error in the previous version. The work is supported by the NSFC Grant Nos. 11890712, 12075129 and 12175165. This study has, furthermore, received funding from the European Union’s Horizon 2020 research and innovation program under grant agreement No 824093 (STRONG-2020).
\bibliographystyle{apsrev4-1.bst}
\bibliography{Ref}
\end{document}